\documentclass[11pt]{article}
\usepackage{moriond,epsfig}

\def\Journal#1#2#3#4{{#1} {\bf #2}, #3 (#4)}

\def\NIMA{{\em Nucl. Instrum. Methods} A}
\def\NPB{{\em Nucl. Phys.} B}

\def\PRL{\em Phys. Rev. Lett.}
\def\PRC{{\em Phys. Rev.} C}

\def\ZPC{{\em Z. Phys.} C}

\newcommand{\snn}{\sqrt{s_{NN}}}
\newcommand{\pbarp}{{\rm p}(\overline{{\rm p}}{\rm )+p}}

\newcommand{\npart}{N_{part}}
\newcommand{\ncoll}{N_{coll}}
\newcommand{\avenp}{\langle\npart\rangle}

\newcommand{\half}{\frac{1}{2}}
\newcommand{\halfnp}{(\half\avenp)}
\newcommand{\nch}{N_{ch}}

\newcommand{\dndeta}{d\nch/d\eta}

\newcommand{\dndetanp}{\dndeta / \halfnp}

\begin{document}
\vspace*{4cm}
\title{BULK OBSERVABLES IN pp, dA AND AA COLLISIONS at RHIC}

\author{DAVID J. HOFMAN$^6$ for the PHOBOS Collaboration \\[0.2cm]
B.B.Back$^1$,
M.D.Baker$^2$, M.Ballintijn$^4$, D.S.Barton$^2$, B.Becker$^2$,
R.R.Betts$^6$, A.A.Bickley$^7$, R.Bindel$^7$, A.Budzanowski$^3$,
W.Busza$^4$, A.Carroll$^2$, M.P.Decowski$^4$, E.Garc\'{\i}a$^6$,
T.Gburek$^3$, N.George$^{1,2}$, K.Gulbrandsen$^4$, S.Gushue$^2$,
C.Halliwell$^6$, J.Hamblen$^8$, A.S.Harrington$^8$,
G.A.Heintzelman$^2$, C.Henderson$^4$, 
R.S.Hollis$^6$, R.Ho\l y\'{n}ski$^3$, B.Holzman$^2$,
A.Iordanova$^6$, E.Johnson$^8$, J.L.Kane$^4$, J.Katzy$^{4,6}$,
N.Khan$^8$, W.Kucewicz$^6$, P.Kulinich$^4$, C.M.Kuo$^5$,
J.W.Lee$^4$, W.T.Lin$^5$, S.Manly$^8$, D.McLeod$^6$,
A.C.Mignerey$^7$, R.Nouicer$^{2,6}$, A.Olszewski$^3$, R.Pak$^2$,
I.C.Park$^8$, H.Pernegger$^4$, C.Reed$^4$, L.P.Remsberg$^2$,
M.Reuter$^6$, C.Roland$^4$, G.Roland$^4$, L.Rosenberg$^4$,
J.Sagerer$^6$, P.Sarin$^4$, P.Sawicki$^3$, I.Sedykh$^2$,
W.Skulski$^8$, C.E.Smith$^6$, P.Steinberg$^2$, G.S.F.Stephans$^4$,
A.Sukhanov$^2$, J.-L.Tang$^5$, M.B.Tonjes$^7$, A.Trzupek$^3$,
C.Vale$^4$, G.J.van~Nieuwenhuizen$^4$, R.Verdier$^4$,
G.I.Veres$^4$, F.L.H.Wolfs$^8$, B.Wosiek$^3$, K.Wo\'{z}niak$^3$,
A.H.Wuosmaa$^1$, B.Wys\l ouch$^4$, J.Zhang$^4$\\[0.2cm]
}

\address{
$^1$Argonne National Laboratory, Argonne, IL 60439-4843, USA\\
$^2$Brookhaven National Laboratory, Bldg. 555, P.O. Box 5000, Upton, NY
11973-5000, USA.\\
$^3$Institute of Nuclear Physics PAN, Krak\'{o}w, Poland\\
$^4$Massachusetts Institute of Technology, Cambridge, MA 02139-4307,
USA\\
$^5$National Central University, Chung-Li, Taiwan\\
$^6$University of Illinois at Chicago, Chicago, IL 60607-7059, USA\\
$^7$University of Maryland, College Park, MD 20742, USA\\
$^8$University of Rochester, Rochester, NY 14627, USA
}

\maketitle\abstracts{
Results on charged particle production in p+p, d+Au and
Au+Au collisions at RHIC energies ($\snn$ = 19.6 to 200 GeV) are 
presented. 
The data exhibit remarkable, and simple, scaling behaviors, 
the most prominent of which are discussed.
}

\section{Introduction}

In this paper, we report on the bulk features of particle 
production, focusing on 
results from the PHOBOS experiment at the Relativistic 
Heavy-Ion Collider (RHIC).  A primary strength of 
PHOBOS is the ability to measure the total charged multiplicity,
with practically no extrapolations necessary, due to the
$\sim 4 \pi$ acceptance of the detector.  The specific detectors 
that allow the measurement of charged particle multiplicity 
($\dndeta$) over the broad pseudorapidity range ($|\eta|<5.4$) 
are the midrapidity Vertex detector ($|\eta|<1$), 
the centrally located Octagon barrel ($|\eta|<3.2$),
and a series of six Ring counters ($3.1<|\eta|<5.4$), all of which
closely surround a thin Be beam-pipe. These detectors
are silicon pad sensors, 
see Ref.~\cite{PHOBOS_NIM} for details.

\section{Bulk Charged Particle Production}
\label{bulk_production}


\begin{figure} [t]
\hspace{0.3cm}
\begin{minipage}{7.2cm}
\psfig{figure=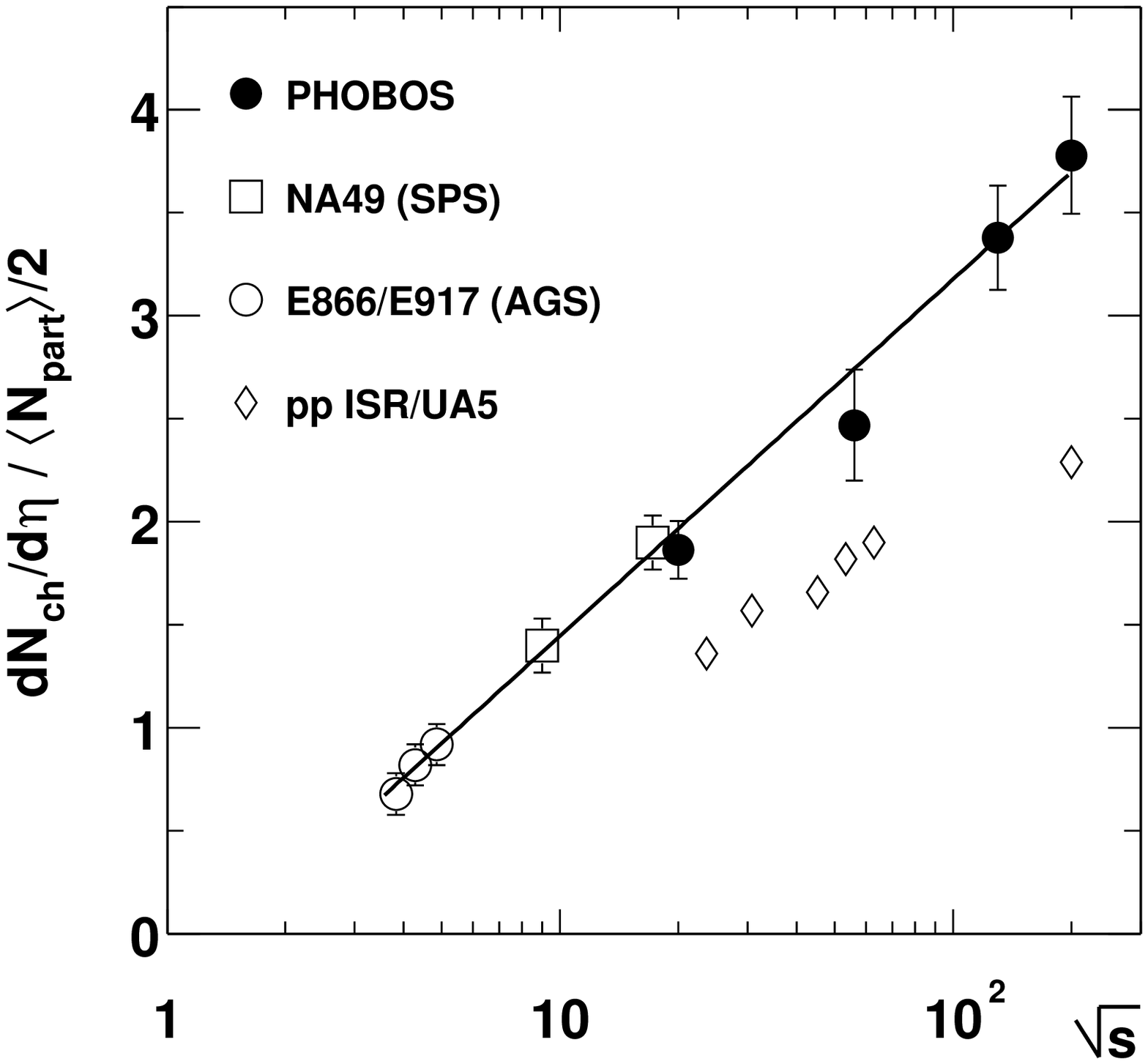,width=6.5cm}\
\caption{Excitation function of midrapidity ($|\eta|<1$) 
charged particle yields per participant pair, $\dndetanp$, from PHOBOS
at RHIC energies for the 6\% most central Au+Au collisions (solid circles).  
Also shown is heavy-ion
data from lower energies and, for a baseline comparison, the midrapidity
($|\eta|<1$) yield of charged particles from 
inelastic $\pbarp$ collisions (open diamonds).}
\label{figure1}
\end{minipage}
\hspace{0.3cm}
\begin{minipage}{7.2cm}
\psfig{figure=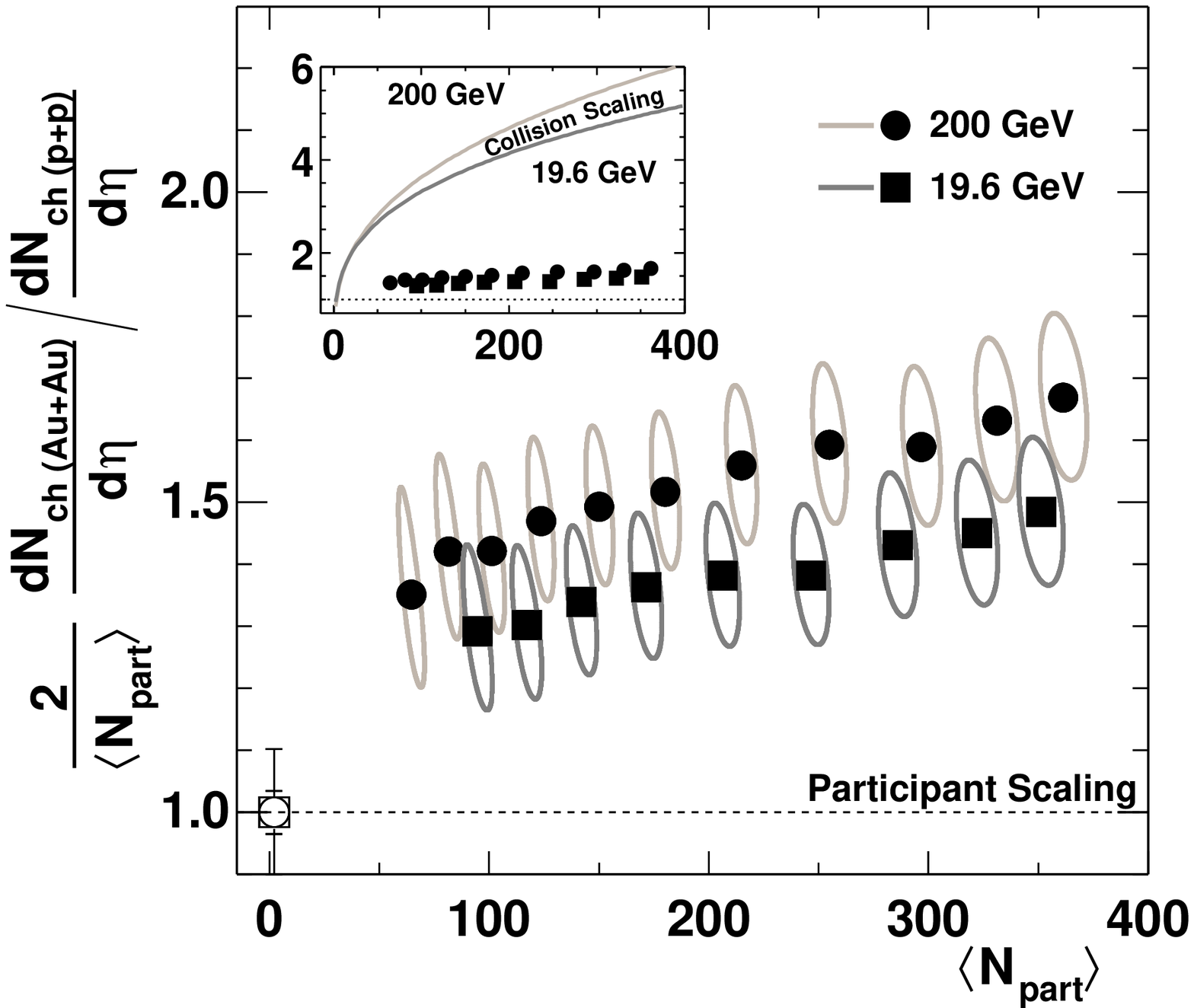,width=6.5cm}
\caption{Centrality, or $\avenp$, dependence of the 
charged particle multiplicity
per participant pair at midrapidity for Au+Au collisions at 
200 and 19.6 GeV divided by the corresponding 
value obtained in inelastic $\pbarp$ collisions.
The inset has the same axis labels as the main figure, with curved lines
representing expectation from pure binary collision ($\ncoll$)
scaling and the horizontal dashed line corresponding
to pure participant ($\npart$) scaling.}
\label{figure2}
\end{minipage}
\end{figure}

A surprising discovery at RHIC is the existence of simple
scaling behaviors observed in bulk charged particle production
(i.e. integrated over p$_T$ and particle species). 
Some of these features are similar to those seen at lower energies, 
as well as in simpler systems, including $\pbarp$ 
and e$^+$+e$^-$ collisions.  These scaling properties are 
found to exist both as a function of the collision 
geometry as well as in the rest-frame of one of the colliding nuclei. 
In the following discussion, it is important to note that the measured
bulk charged particle multiplicities are completely dominated by 
the emission of low p$_T$ ($\leq$ 1.5 GeV/c) particles.

Determining the centrality (or impact parameter $b$) 
of a heavy-ion collision is
extremely important to provide a geometrical scale for use 
in the studies of underlying collision dynamics.
The event centrality is characterized by charged particle multiplicities
measured in various regions of phase-space.  Comprehensive
Monte Carlo (MC) simulations of these signals, that include 
Glauber model calculations of the collision geometry, allow the
estimation of $\npart$, the number of participating nucleons 
in the collision~\cite{PHOBOS_130_200_cent,PHOBOS_200_20_ratio},
for a selected class of events.
The most central ($b \sim 0$) collisions will have the largest number
of participants with the obvious upper limit of 394 for a ``perfectly 
central'' Au+Au collision where all the nucleons interact.  
For a particular collision geometry, the MC simulation also allows 
calculation of the number of binary collisions, $\ncoll$, which provides
an expected baseline scaling for large-momentum transfer processes.

One of the first measurements at RHIC was the multiplicity of 
charged particles, $\dndeta$ (pseudorapidity density), 
produced in head-on collisions of 
Au+Au nuclei and emitted at midrapidity, defined here as $|\eta|<1$. 
This and subsequent results~\cite{PHOBOS_200_20_ratio,PHOBOS_midrap} 
are shown in Fig.~\ref{figure1}, where the midrapidity $\dndeta$ 
is divided by the average number of participating nucleon pairs,
$\halfnp$, in order to compare directly to results from
$\pbarp$ collisions.
What is most notable in the heavy-ion data of Fig.~\ref{figure1} is 
the apparent linear logarithmic growth of the midrapidity particle 
production per participant pair with collision energy ($\snn$). 
This multiplicity in central Au+Au collisions
is larger, however, than seen in inelastic $\pbarp$
collisions at similar energy, indicated by the open 
diamonds~\cite{ISRppUA5} in Fig.~\ref{figure1}.


\begin{figure} [t]
\hspace{0.3cm}
\begin{minipage}{7.2cm}
\psfig{figure=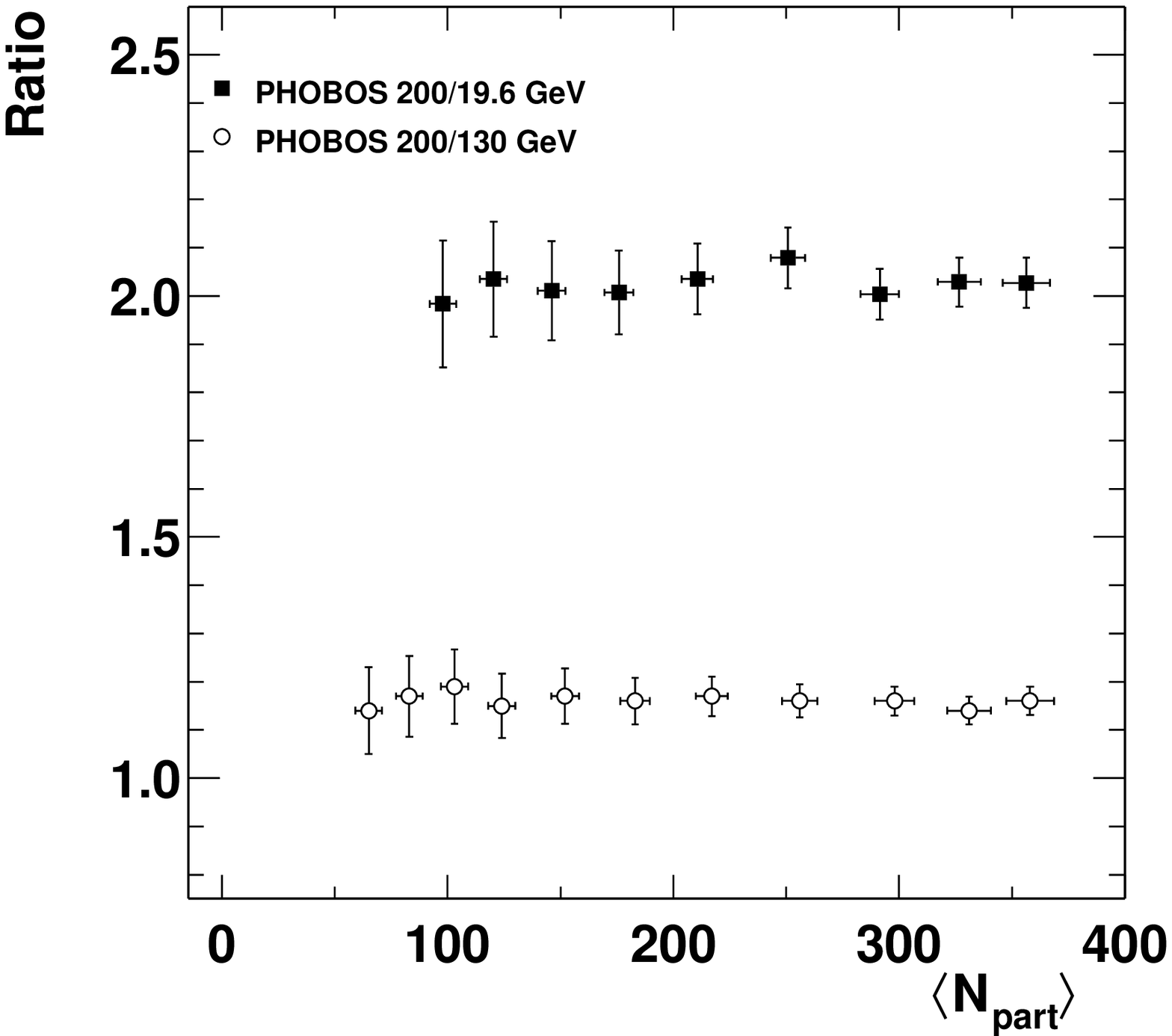,width=6.5cm}\
\caption{Centrality, or $\avenp$, dependence of the 
ratio of midrapidity charged particle multiplicity per 
participant pair for Au+Au collisions measured at $\snn$ = 200 
GeV to that obtained at lower energies $\snn$ = 130 and 19.6 GeV.}
\label{figure3}
\end{minipage}
\hspace{0.3cm}
\begin{minipage}{7.2cm}
\psfig{figure=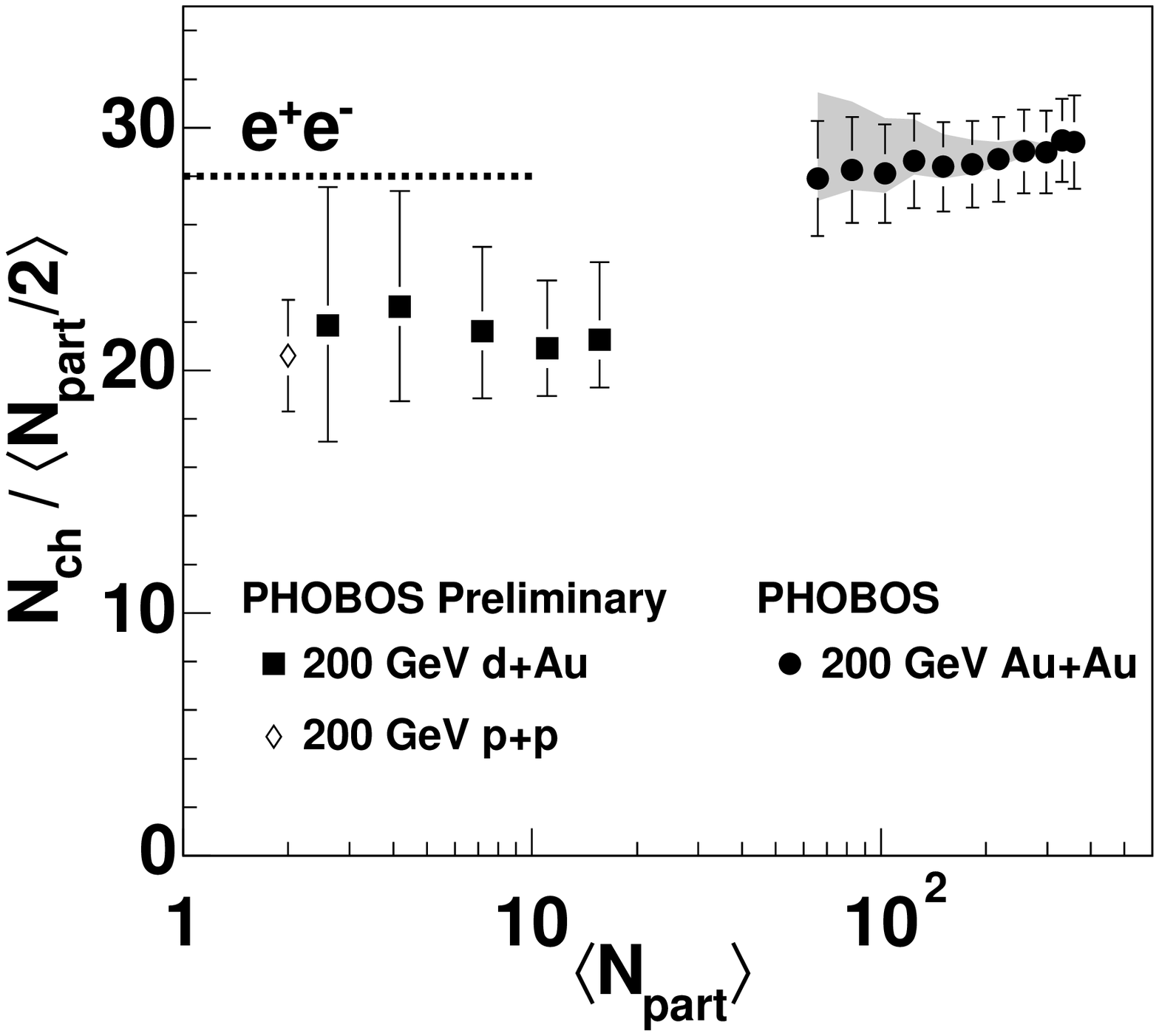,width=6.5cm}
\caption{Comparison of total charged particle 
multiplicities per participant pair in p+p, d+Au and 
Au+Au collisions versus centrality at $\snn$ = 200 GeV.  
Also shown (dashed line) 
is the value from elementary e$^+$+e$^-$ collisions.}
\label{figure4}
\end{minipage}
\end{figure}

The centrality dependence of midrapidity multiplicity per participant pair
in Au+Au collisions at $\snn$ = 200 and 19.6 GeV is shown 
in Fig.~\ref{figure2}, where the heavy-ion data has been 
normalized by the corresponding multiplicity in inelastic $\pbarp$ 
collisions. 
The most dramatic feature of the Au+Au centrality dependence
is the similarity of the collision geometry scaling
at both energies, a scaling which follows a strong $\npart$-like 
dependence.
Despite a factor of 10 increase in collision
energy from $\snn$ = 19.6 to 200 GeV, 
both data sets are consistent with 
only a small ($\approx$ 13\%) fraction of binary collision
($N_{coll}$) scaling~\cite{PHOBOS_200_20_ratio}.

Although there is a slight increase in midrapidity multiplicity
per participant pair in Au+Au collisions as a function of 
centrality (see Fig.~\ref{figure2}), we find this increase exhibits a
simple scaling at RHIC. 
This scaling is shown in Fig.~\ref{figure3}, where the 
ratio of 200/130 and 200/19.6
GeV midrapidity $\dndetanp$ is given as function of centrality.  Within 
errors, the ratio is independent of centrality for the most
central 40\% of the cross section
and yields a simple scale factor
R$_{200/130}$ = 1.14$\pm$0.01(stat)$\pm$0.05(syst)~\cite{PHOBOS_130_200_cent} 
and R$_{200/19.6}$ 
= 2.03$\pm$0.02(stat)$\pm$0.05(syst)~\cite{PHOBOS_200_20_ratio}.

Expanding to the full phase space, we find the midrapidity increase in
$\dndetanp$ with centrality
is remarkably compensated by a corresponding decrease in yield
at high $\eta$ such that the total charged particle production 
per participant pair is independent of centrality.  
This can be seen in Fig.~\ref{figure4} for both Au+Au collisions 
as well as for d+Au collisions.  We also observe that the Au+Au 
and d+Au data give different values of $\nch/\halfnp$, 
but they are different in a very interesting way.
For the same collision energy, the Au+Au data matches 
that seen in e$^+$+e$^-$ collisions and the d+Au data matches 
that seen in $\pbarp$ collisions.  


\begin{figure} [t]
\hspace{0.3cm}
\begin{minipage}{7.2cm}
\psfig{figure=fig5_moriond.eps,width=6.5cm}\
\caption{$\dndeta$ as function of $\eta - y_{beam}$
for $\pbarp$ data from the ISR and
UA5 {\protect \cite{ISRppUA5}}.}
\label{figure5}
\end{minipage}
\hspace{0.3cm}
\begin{minipage}{7.2cm}
\psfig{figure=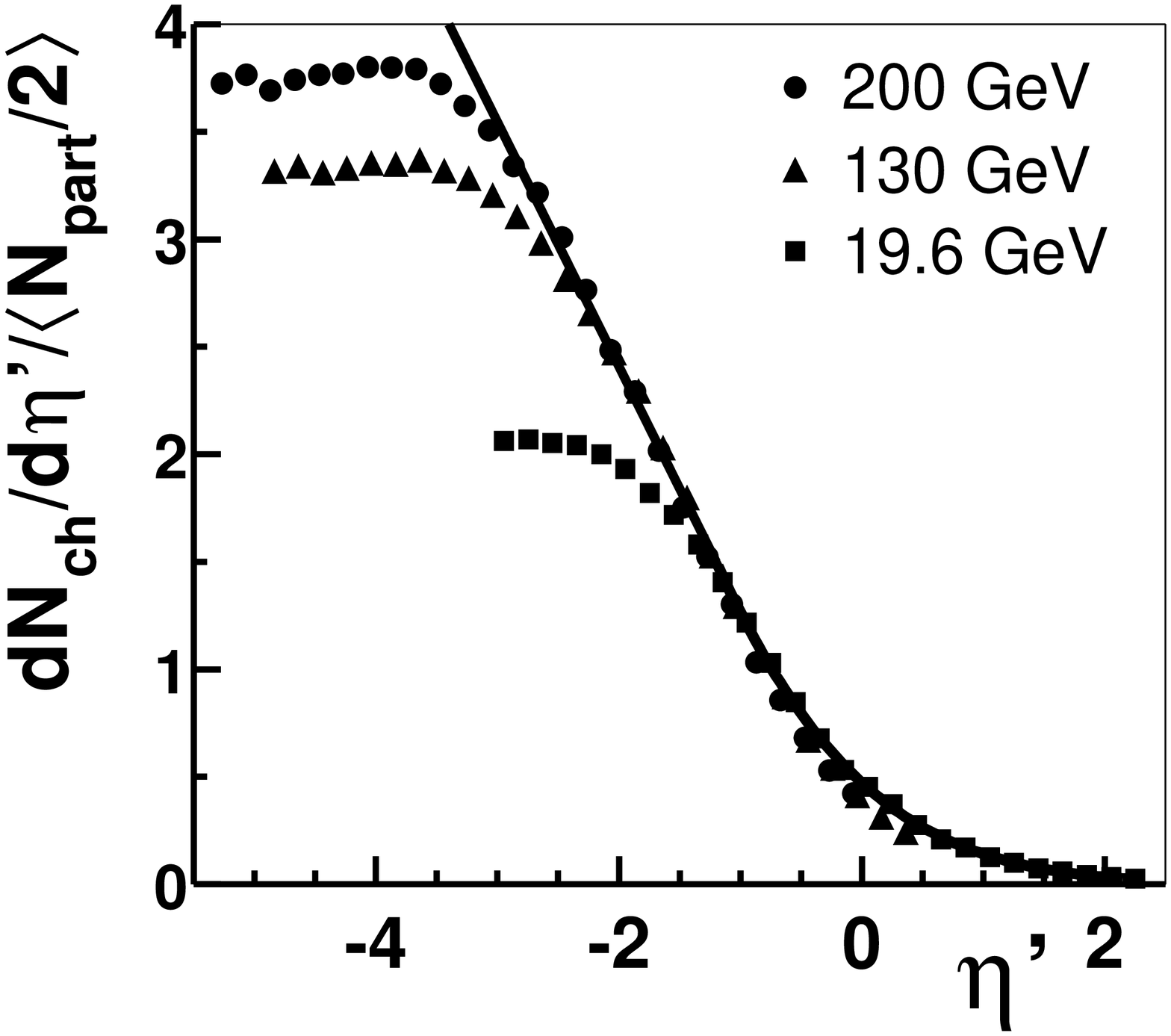,width=6.5cm}
\caption{$\dndeta'/\halfnp$ as a function of $\eta'$ for Au+Au data 
 from PHOBOS at RHIC {\protect \cite{PHOBOS_limitfrag}}.}
\label{figure6}
\end{minipage}
\end{figure}

The last scaling feature we examine is evident when particle
production is viewed in the rest frame of one of the colliding nuclei.  
This is approximately achieved by effectively transforming 
measured $\dndeta$ multiplicities 
to the rest frame of the beam with rapidity $y_{beam}$, using
$dN/d\eta'$ where $\eta' = |\eta| - y_{beam}$. 
In the late 1960's it was hypothesized that in hadronic collisions, 
at high enough collision energy, the charged particle yield 
(d$^2$N/dy$'$dp$_T$) reaches a limiting value and becomes 
independent of energy when viewed near 
beam rapidity~\cite{Benecke69}. This ``limiting fragmentation'' 
hypothesis works in many elementary systems including 
$\pbarp$ as shown in the $\dndeta$ multiplicities 
in Fig.~\ref{figure5}.  We find the same
features in the Au+Au heavy-ion data at RHIC~\cite{PHOBOS_limitfrag} 
(see Fig.~\ref{figure6}).  The same ``limiting fragmentation'' features
are also seen in $\dndeta$ data from 
d+Au collisions~\cite{Nouicer_qm2004}.   
Most recently, PHOBOS has additionally observed the same type of scaling
in the magnitude of elliptic flow, $v_2(\eta')$, for heavy-ion 
collisions~\cite{PHOBOS_v2limitfrag}.

\section{Conclusions}

We report here a few of the bulk properties of charged particle production
seen in relativistic heavy-ion collisions at RHIC energies.  We find
that much of the data can be described in terms of simple scaling
behaviors.  These observations are not 
yet understood, but could suggest the existence of strong 
global constraints or some form of universality in
the bulk hadron production mechanism.  These 
behaviors must be understood before a complete picture 
of the properties of the high energy-density matter created 
at RHIC can be definitively determined.

\section*{Acknowledgments}
This work was partially supported by U.S. DOE grants
DE-AC02-98CH10886, DE-FG02-93ER\-40802,
DE-FC02-94ER40818,  
DE-FG02-94ER40865, DE-FG02-99ER41099, and W-31-109-ENG-38, by U.S.
NSF grants 9603486, 
9722606,
0072204,            
and 0245011,        
by Polish KBN grant 2-P03B-10323, and by NSC of Taiwan Contract
NSC 89-2112-M-008-024.


\section*{References}

\newcommand{\etal} {$\mathrm{\it et\ al.}$}


\begin{thebibliography}{99}

\bibitem{PHOBOS_NIM} B.B.~Back \etal\ (PHOBOS), 
	\Journal{\NIMA}{499}{603}{2003}

\bibitem{PHOBOS_130_200_cent} B.B.~Back \etal\ (PHOBOS), 
	\Journal{\PRC}{65}{061901(R)}{2002}

\bibitem{PHOBOS_200_20_ratio} B.B.~Back \etal\ (PHOBOS), 
	arXiv:nucl-ex/0405027

\bibitem{PHOBOS_midrap} B.B.~Back \etal\ (PHOBOS), 
	\Journal{\PRL}{88}{022302}{2002}

\bibitem{ISRppUA5} W.~Thom\'{e} \etal\ (CERN-ISR), 
	\Journal{\NPB}{129}{365}{1977};
	G.J.~Alner \etal\ (UA5), \Journal{\ZPC}{33}{1}{1986}; and
        R.E.~Ansorge \etal\ (UA5), \Journal{\ZPC}{43}{357}{1989}

\bibitem{Benecke69} J.~Benecke \etal, 
	\Journal{Phys. Rev}{188}{2159}{1969}

\bibitem{PHOBOS_limitfrag} B.B.~Back \etal\ (PHOBOS), 
	\Journal{\PRL}{91}{052303}{2003}

\bibitem{Nouicer_qm2004} R.~Nouicer \etal\ (PHOBOS), 
	arXiv:nucl-ex/0403033

\bibitem{PHOBOS_v2limitfrag} B.B.~Back \etal\ (PHOBOS), 
	arXiv:nucl-ex/0406021

\end{thebibliography}
\end{document}